%
%
%
%
%
\input iopppt
\pptstyle
\jl{1}
\letter{New non-unitary representations in a Dirac hydro- gen atom,}
%
%
 
 \def\ie{{\it i.e.\ }}
 
%
 \newbox\Ancha
 \def\gros#1{{\setbox\Ancha=\hbox{$#1$}
   \kern-.025em\copy\Ancha\kern-\wd\Ancha
   \kern.05em\copy\Ancha\kern-\wd\Ancha
   \kern-.025em\raise.0433em\box\Ancha}}
%

\author{ R P Mart\'inez-y-Romero, J Salda\~na-Vega and A L Salas-Brito\footnote{\dag}{On sabbatical leave from Laboratorio de Sistemas Din\'amicos, UAM-Azcapotzalco, e-mail: asb@data.net.mx or asb@hp9000a1.uam.mx} }[R P Mart\'inez-y-Romero \etal ]

\address{  Facultad de Ciencias, Universidad Nacional Aut\'onoma de M\'exico, Apartado Postal 21-726, C P 04000, Coyoac\'an  D F, M\'exico}

\abs
   New non-unitary representations of the SU(2) algebra are introduced for the case of the Dirac equation with a Coulomb potential; an extra phase, needed to close the algebra, is also introduced. The new representations does not require integer or half integer labels. The  set of operators defined are used to span the complete space of bound state eigenstates of the problem thus solving it in an essentially algebraic way.
\endabs

\submitted 

\date


 Hydrogen-like atoms are of the most important quantum systems solved. Even for describing stabilization properties  and  for testing QED and weak interaction theories a great deal can be done at the relativistic atomic physics level (Greiner 1991, Kylstra \etal\ 1997, Quiney \etal\ 1997). It is very important then any insight that might be given on the properties of hydrogen-like systems. An important tool have been the algebraic properties of the set of operators defining the system; these are not only connected with the corresponding group and its symmetry algebra but often offer also simplified methods for doing some calculations. It is the purpose of this Letter to define a new set of operators for the Dirac relativistic hydrogen atom. This comprises a non unitary representation of the SU(2) algebra and defines ladder operators for the problem. An extra phase is needed to close the algebra but this allows us to solve the Dirac hydrogen atom in a neat algebraic way. \par

The Dirac Hamiltonian for an hydrogen-like atom is

$$ H= {\gros\alpha}\cdot {\bf p} + \beta m - {Z e^2\over r},   \eqno(1)  $$

\noindent where $\gros\alpha$ and $\beta$ are Dirac matrices (Bjorken and Drell 1964), $Z$ is the atomic number, $r$ is the relative distance between the electron and the nucleus, $m$ is the  mass, and we use units such that $\hbar=c=1$. \par

Taking advantage of the obvious rotational symmetry of $H$, we can express the bound eigenstates of the hydrogen atom as

$$ \psi ({\bf r}, t) = {1 \over r} \pmatrix { F(r) {\cal Y}_{jm}(\theta, 
\phi) \cr  \i G(r) {\cal Y}\,'_{jm}(\theta, \phi) }, 
\eqno(2) $$ 

\noindent where ${\cal Y}_{jm}$ and ${\cal Y}\,'_{jm}$ are spinor spherical harmonics of opposite parity and $j$ is the total angular momentum (Greiner 1991).  It is convenient to define the quantum number $\epsilon$ such that it equals $+1$ when $l=j+1/2$ and it equals $-1$ when $l=j-1/2$ and use it instead of parity. The `big' and `small' radial components of the  bi-spinor (2) describing bound states of an hydrogen atom are solutions of the system

$$ \left( -{\d \over \d\rho } + {\tau_j \over \rho} \right) G(\rho)
= \left(-\nu + {\zeta\over \rho} \right) F(\rho), \eqno(3) $$

$$ \left( +{\d \over \d\rho } + {\tau_j \over \rho} \right) F(\rho)
= \left(\nu^{-1} + {\zeta\over \rho} \right) G(\rho) \eqno(4) $$

\noindent where we have defined the positive definite quantity $k:=\sqrt{m^2-E^2}$, $E$ being the energy of the bound state and we have expressed the equations in terms of the dimensionless variable $\rho:= kr$. For the sake of simplicity we also used $\zeta:= Ze^2$, $\tau_j:= \epsilon(j+1/2)$, and $\nu:= \sqrt{(m-E)/(m+E)}$.\par

 Let us now change to the new variable $x$ defined by $\rho=\e^x$ so that the range of $x$ is the open interval $(-\infty, \infty)$ and redefine the radial functions as

$$ 
\eqalign{ F(\rho(x)):=& \sqrt{m+E} \left[  \psi_-(x) + \psi_+(x) \right],  \cr
          G(\rho(x)):=& \sqrt{m-E} \left[  \psi_-(x) - \psi_+(x) \right]. } \eqno(5)
$$

\noindent With the new functions $ \psi_+(x) $ and $ \psi_-(x)$, defining

$$ \mu := {\zeta E\over k}+1 \eqno(6) $$

\noindent and after some manipulations we get the following system of differential equations

$$  
\left[{\d^2 \over \d x^2} + 2\mu \e^x - \e^{2x}-{1\over 4} \right]\psi_+(x) = \left( \tau_j^2-\zeta^2-{1\over 4}  \right) \psi_+(x), \eqno(7)\cr
\left[{\d^2 \over \d x^2} + 2(\mu-1) \e^x - \e^{2x}-{1\over 4} \right]\psi_-(x) = \left( \tau_j^2-\zeta^2-{1\over 4}  \right) \psi_-(x), \eqno(8)
$$

\noindent  for describing an the radial part of an hydrogen atom. As it should become clear in what follows, the inclusion of the term 1/4 in the above equations is necessary to close the algebra we purport to construct. Notice that equations (7) and (8) can be regarded as  an eigensystem with  the known eigenvalue 

$$\omega:= \tau_j^2-\zeta-1/4 =j(j+1)-\zeta^2, \eqno(9) $$

\noindent as follows  from the radial symmetry of the hydrogen atom.  \par

In order to rewrite the system (7) and (8) making clear the relation with a SU(2) algebra let us  define the two operators

$$ \Omega_\pm:=  \e^{\pm\i\xi}\left({\partial \over\partial x} \mp \e^x
\mp\i{\partial\over \partial\xi} + {1\over 2} \right), \eqno(10) $$

\noindent where we introduced the extra phase $\xi$ besides the `radial' variable $x$, and a third operator

$$ \Omega_3 := -\i {\partial \over \partial \xi}, \eqno(11)  $$

\noindent which depends exclusively on $\xi$; we can alternatively define the two operators $\Omega_1$ and $\Omega_2$ as

$$ \Omega_1= {1\over 2} \left(\Omega_+ + \Omega_-\right), \qquad \Omega_2= {1\over 2i} \left(\Omega_+ - \Omega_-\right). \eqno(12) $$

The previously defined operators are easily seen to satisfy all the relationships of the SU(2) algebra

$$ [\Omega_i, \Omega_j]=\i \epsilon_{ijk} \Omega_k, \quad i,j,k =1,2,3, \eqno(13) $$

\noindent where Einstein summation convention is implied; the algebra can be also expressed in terms of the operators $\Omega_\pm$

$$ [\Omega_3, \Omega_\pm] = \pm \Omega_\pm, \quad \hbox{ and }\quad
 [\Omega_+, \Omega_-]= 2 \Omega_3,        \eqno(14) $$

\noindent which thus play the role of raising and lowering operators.  The operator

$$ \Omega^2:={\gros\Omega}{\bf\cdot}{\gros\Omega}=\Omega_1^2+\Omega_2^2+\Omega_3^2={\partial^2\over \partial x^2}-\e^{2x}-2\i\e^x{\partial \over \partial \xi}-{1\over 4}, \eqno(15) $$

\noindent where ${\gros \Omega}:= \Omega_1 {\bf i}+ \Omega_2{\bf j} +\Omega_3{\bf k}$, is the Casimir operator since $[\Omega^2, \Omega_i]=0$ for all $i=1,2,3$. \par

Given the just stated properties, we can choose the operators $\Omega^2$ and $\Omega_3$ and define the simultaneous eigenstates $V_\omega^\mu(x,\xi)$ where $\omega$ and $\mu$ are, repectively, the eigenvalues of $\Omega^2$ and $\Omega_3$  ---the notation  is chosen in analogy with the spherical harmonic $Y_l^m(\theta, \phi)$ case. It must be keep in mind though that we are not here restricted to a compact set of parameters given the infinite range of $x$. Our choice of eigenstates allows us to write

$$ \Omega_3 V_\omega^\mu(x,\xi)=\mu V_\omega^\mu(x,\xi), \quad \Omega^2 V_\omega^\mu(x,\xi)=\omega V_\omega^\mu(x,\xi), \eqno(16)  $$

\noindent and thus 

$$V_\omega^\mu(x,\xi)= \e^{\i \mu\xi}{\cal P}^\mu_\omega(x), \eqno(17)$$

\noindent where we have tried to use a notation for the $x$-function reminiscent of the associated Legendre polynomials. 

Now, using equations (7) and (8) we can straightway  get  $\psi_+(x)= V_\omega^\mu(x,\xi)$ and $\psi_-(x)=V_\omega^{\mu-1}(x,\xi)$; besides, we can easily show that the operators $\Omega_\pm$ change the eigenvalue $\mu$ to the eigenvalue $\mu\pm1$, \ie\ 

$$\Omega_\pm V_\omega^\mu(x,\xi) =C_\mu^{\pm} V_\omega^{\mu\pm1}(x,\xi), \eqno(18)$$

\noindent where the numbers $C_\mu^{\pm}$ can be determined from $<\omega\mu|\Omega_+\Omega_-|\omega\mu>=C_\mu^-C_{\mu-1}^+$ and, with an appropriate selection of phase, they become $C_\mu^{\pm}=\pm\sqrt{\mu(\mu\pm1)-\lambda(\lambda-1)}$. These definitions also establish the conection of our operators with the hydrogen atom problem since applying $\Omega^2$ to $V_\omega^\mu(x,\xi)$ or to $V_\omega^{\mu-1}(x,\xi)$ essentially reproduces equations (7) and (8) ---as the second of equations (16) clearly exhibits.

For establishing the Hermiticity (or lack thereof) of the operators introduced, we need a scalar product; to this end it suffices to use the following product

$$ <\phi, \psi>=\int_0^{2\pi} {d\xi\over 2\pi} \int_{-\infty}^{\infty} \phi^*(\xi,x)\psi(\xi,x)\, dx. \eqno(19) $$ 

\noindent With this interior product, the eigenstates $V_\omega^{\mu}(x,\xi):=|\omega, \mu>$ form a complete orthogonal basis $<\omega', \mu'|\omega,\mu>=\delta_{\mu\mu'}\delta_{\omega\omega'}$. Definition (19) also implies  that $\Omega_3$ is an Hermitian operator, but that $\Omega_1=-\Omega_1^\dagger$ and $\Omega_2=-\Omega_2^\dagger$, that is they are anti-Hermitian (Mart\'inez-y-Romero \etal\ 1997). Thence $\Omega^2$ is not necessarily positive definite; a positive definite operator can be anyway defined as

$$ {\gros\Omega}^\dagger{\bf\cdot}{\gros\Omega}=\Omega_3^2-\Omega_1^2-\Omega_2^2=2\Omega_3^2-\Omega^2. \eqno(20) $$

\noindent The action of this operator can be shown to imply that $2\mu^2\ge \omega$, meaning that $|\mu|$ is bounded by below; let us call $\lambda$ this minimum value \ie\ $\lambda:=|\mu|_{\hbox{min}}$. With this we easily get that $\omega=\lambda(\lambda-1)$ and so, since $\omega=\tau_j^2-\zeta^2-1/4$ and we are looking for positive $\lambda$,

$$ |\mu|_{\hbox{min}}=\lambda=s+{1\over 2}, \eqno(21) $$

\noindent where  $s:=+(\tau_j^2-\zeta^2)^{1/2}$. The most important conclusion we can draw from our discussion is that $\lambda$  no longer has to be restricted to integer or half-integer values as happens necessarily in the standard angular momentum or SU(2)  (Hermitian) case. \par

This curious and interesting  result means that looking for solutions to the Dirac hydrogen atom can be regarded also as looking for non-unitary representations of a SU(2) algebra labeled by real numbers $\lambda$ ---that is, no longer restricted to integer or half-integer numbers. In fact, according to equation (21) we have two series of eigenvalues depending on whether $\mu$ is a positive number or not. In the first case, $\lambda$ is the  minimum number of an infinite set of positive eigenvalues: $ \mu=-\lambda-k$, in the second case, $-\lambda$ is the maximum value of the infinite set of negative eigenvalues: $\mu=-\lambda-k$; in both cases  $k=0,1,2,\dots$ is a non-negative integer. Notice that from the physical point of view the existence of infinite representations of the SU(2) algebra makes sense since it is associated with the denumerably infinite  set of energy eigenvalues of the hydrogen atom. In fact, the energy spectrum of the system is easily obtained from (6) and is found to be

$$E=m\left[ 1 + {\zeta^2\over (\mu-1/2)^2}\right]^{-1/2};    \eqno(22)$$

\noindent in the case of positive eigenvalues we have $\mu=\lambda+k=s+k+1/2$, $k=0,1,2,\dots$, this is precisely the energy spectrum for an Dirac hydrogen atom (Bjorken and Drell 1964, Greiner 1991). The negative eigenvalues do not led to physically admissible eigenstates as we shall see in what follows. \par

 The ground state of the system follows from the equation $\Omega_-|\lambda\lambda>=0$ for the positive eigenvalues. The solution of this differential equation can be found to be

$$ {\cal P}_\lambda^\lambda(x)={2^{(\lambda-1/2)}\over \sqrt{\Gamma(2\lambda-1)}}\e^{(\lambda-1/2)x}
\exp(-\e^x), \eqno(23)$$

\noindent where $\Gamma(y)$ stands for the Euler-gamma function. As $\lambda$ is the lowest eigenvalue, we trivially obtain

$$ \psi_+(x)= {\cal P}_\lambda^\lambda(x),\quad\hbox{and}\quad \psi_-(x)= 0; \eqno(24) $$

\noindent thus the base state is given by $|\lambda \lambda>= \e^{\i \lambda \zeta} {\cal P}_\lambda^\lambda$. Using the original variable $\rho$, the big and the small components for the ground state can be shown to behave as $F(\rho) \propto \sqrt{m+E}\rho^s\e^{-\rho}$ and $G(\rho) \propto \sqrt{m-E}\rho^s\e^{-\rho}$ and, in the negative eigenvalue case, the solution behaves as $\sim \rho^s \e^\rho$ resulting in a divergent behaviour as $\rho\to \infty$  making it unsuitable as an eigenfunction of the hydrogen atom; it is obvious that every other negative eigenvalue function is also unsuitable. \par

The excited states are obtained applying $\Omega_+$ succesively to $|\lambda \lambda>$; they end being polynomials multiplied by the weight factor $\rho^{\lambda-1/2}\e^{-\rho}$ which assures the appropriate behaviour of the eigenstates both as $\rho\to 0$ as well as $\rho\to\infty$. A more detailed discussion of these new polynomials and graphical representations of them are given in a more detailed paper (Mart\'inez-y-Romero \etal  1997). \par

 The main conclusion of this Letter has to do with the infinite-dimensional non-unitary representation of SU(2)  and where each of the basic operators, excepting $\Omega_3$, are also non-Hermitian. For example, the matrix elements of $\Omega_3$  are

$$ <\omega\mu|\Omega_3|\omega\mu'>=\mu\delta_{\mu\mu'},     \eqno(25)  $$

\noindent where $\mu=\pm(\lambda+k)$, $k=0,1,2,\dots$, so its trace vanishes and the determinant of an element of the group of the form $\exp(\i\Omega_3)\xi $ is always 1. The other two operators have as their only non-zero matrix elements

$$ <\omega\mu|\Omega_1|\omega\mu\pm1>= \mp{1\over 2}\sqrt{\mu(\mu\pm1)-\lambda(\lambda-1)}    \eqno(26)  $$

\noindent and

$$ <\omega\mu|\Omega_2|\omega\mu\pm1>= -{\i\over 2}\sqrt{\mu(\mu\pm1)-\lambda(\lambda-1)}.    \eqno(26)  $$

\noindent This means that the trace of both $\Omega_1$ and $\Omega_2$ vanish and that the determinant of group elements is  one only when the  parameter takes imaginary values. We notice also that $\Omega_3$ can be given the physical interpretation of producing infinitesimal changes in the phase of the state $|\lambda \omega>=\e^{\i\mu\xi}{\cal P}_\omega^\lambda(x)$; that is, it can be associated with  the unitary operator $U=\e^{\i\Omega_3\xi}$ which  changes the phase of any state. \par

In summary, we have constructed an SU(2) algebra for the relativistic  hydrogen in the Dirac formulation where the corresponding group is not necessarily compact. We must pinpoint that in order to close the algebra we were forced to introduce an extra parameter $\xi$ which plays the role of a phase.  One of the most noteworthy features of the representations reported here is the mixing of a  spinorial angular momentum character (implying an equally spaced spectrum) with the energy requirements of the problem (requiring a differently spaced spectrum); the interplay of these two spectral requirements is basically reflected in equations (15), (20), (21) and (22), and in the fact that the eigenvalues of the system (7), (8) follows from both the generic radial symmetry and the specific features of the interaction.  From equation (6) we may also notice that in the limit of vanishing interaction, $\zeta\to 0$, the representation collapses and, in this case, $\mu=1$ always. This behaviour is precisely as expected because there is no longer any restriction over the eigenvalues and thus the spectrum becomes continuous ---corresponding to a free Dirac particle.

   The representations of the algebra are labeled by numbers which are neither integers nor half-integers as ought to be in the case of the ordinary unitary representations. Nevertheless, the operator algebra introduced allows an essentially algebraic solutiond of Dirac hydrogen atom which may have various applications (De Lange and Raab 1991). It is to be noted also the similitude of our transformed equations with the corresponding ones for the case of the Morse potential (N\'u\~nez-Y\'epez \etal 1995, 1997) and the conections that all our formulation has with systems with hidden supersymmetric properties (Ben\'itez \etal 1990a,b,   Mart\'inez-y-Romero \etal 1991, Mart\'inez-y-Romero and Salas-Brito 1992, Haag \etal 1976) as we will discuss at lenght in a forthcoming article. \par

\ack This work has benefitted of  the comments of R Jauregui, L F Urrutia, A Gonz\'alez-Villanueva and H N N\'u\~nez-Y\'epez and has been partially supported by CONACyT (grant 1343P-E9607). ALSB also acknowledges the friendly support of R Micifuz, P A Terek, Q Gris, F Sadi, B Kot, B Caro, and U Sasi. \par

\references
\refjl {Ben\'itez J, Mart\'inez-y-Romero R P, N\'u\~nez-Y\'epez H N and Salas-Brito A L 1990} 
{Phys. Rev. Lett.} { 64} {1643}
 \refjl {\dash 1990}{Phys. Rev. Lett.}  {65} {2085(E) }

\refbk{Bjorken  J D and Drell S D 1964} {\it Relativistic Quantum Mechanics} {(New York: Mac Graw-Hill)}
\refbk{De Lange O L and Raab R E 1991} {\it Operator Methods in Quantum Mechanics} {(Oxford: Clarendon Press)}
\refbk{Greiner W 1991} {\it Theoretical Physics 3: Relativistic quantum 
mechanics} {(Berlin: Springer)} 
\refjl{Haag R, Lopuszanski V T and Sohnius M 1976} {Nucl. Phys. B} {88} {383}
\refjl{Kylstra N J, Ermolaev A M and Joachaim C J 1997} {J. Phys. B: At. Mol. Opt. Phys.} {30} {L449}
\refjl {Mart\'inez-y-Romero R P, Salda\~na-Vega J and Salas-Brito A L 1999} {J. Math. Phys.}{ 40}  {2324} 

\refjl {Mart\'inez-y-Romero R P and Salas-Brito A L 1992} { J. Math. Phys. } { 33} {1831}
\refjl {Mart\'inez-y-Romero R P, Moreno M and Zentella A 1991} {Phys. Rev. D} {43} {2306}
\refjl{N\'u\~nez-Y\'epez H N, L\'opez J L and Salas-Brito A L 1995} {J. Phys. B: At. Mol. Opt. Phys.} {28} {L525}
\refjl {N\'u\~nez-Y\'epez H N, L\'opez J L, Navarrete D and Salas-Brito A L 1997} {Int. J. Quantum Chem.} {62} {177} 
\refjl {Quiney H M, Skaane H and Grant I P 1997} {J. Phys. B: At. Mol. Opt. Phys.} {30} {L829}

 \vfill
 \eject
  \end